\documentclass[showpacs]{revtex4}
\def\maj#1{\ifmmode\mbox{\usefont{U}{msb}{m}{n}#1}\else{\usefont{U}{msb}{m}{n}#1}\fi}
\def\v#1{\mathbf{#1}}

\usepackage{graphics,epsfig,graphicx}
\usepackage{color}

\begin{document}

\title{\textbf{Biexciton oscillator srength}}
\author{M. Combescot and O. Betbeder-Matibet
\\ \small{\textit{Institut des NanoSciences de Paris,}}\\
\small{\textit{Universit\'e Pierre et Marie Curie, 
CNRS,}}\\
\small{\textit{Campus Boucicaut, 140 rue de Lourmel, 75015 Paris}}}

\begin{abstract}
Our goal is to provide a physical understanding of the elementary coupling between photon and biexciton and to derive  the physical characteristics  of the biexciton oscillator strength, following the procedure we used for trion. Instead of the more standard two-photon absorption, this work concentrates on molecular biexciton created by photon absorption in an exciton gas. We first determine the appropriate set of coordinates in real and momentum spaces to describe one biexciton as two interacting excitons. We then turn to second quantization and introduce the ``Fourier transform in the exciton sense'' of the biexciton wave function which is the relevant quantity for oscillator strength. We find that, like for trion, the oscillator strength for the formation of one biexciton out of one photon plus a \emph{single} exciton is extremely small: it is one biexciton volume divided by one sample volume smaller than the exciton oscillator strength. However, due to their quantum nature, trion and biexciton have absorption lines which  behave quite differently. Electrons and trions are fermionic particles impossible to pile up all at the same energy. This would make the weak trion line spread with electron density, the peak structure                                                                                                                                                                                                                                                                                                                                                                                          only coming from singular many-body effects. By contrast,  the bosonic nature of exciton and biexciton makes the biexciton peak mainly rise with exciton density, this rise being simply linear if we forget many-body effects between the photocreated exciton and the excitons present in the sample.
\end{abstract}


\maketitle

\section{Introduction}

In addition to their well-known interest in today's technology, semiconductors are quite interesting materials from a purely fundamental point of view because they allow us to study a large variety of quantum objects: two-fermion object like exciton, three-fermion object like trion \cite{T0,T1,T2,T3,T4,T5,T6,T7,T8,T9,T10,T11,T12,T13,T14,T15,T16,T17,T17bis,T18,T19}, four-fermion object like biexciton \cite{B1,B2,B3,B4,B5,B5bis,B6,B7,B7bis,B7ter,B7-4,B7-4bis,B7-5,B7-6,B8,B9,B10,B11,B11bis,B11ter,B11-4,B12,B13,
B14,B15,B15bis,B16,B16bis,B16ter,B17,B18,B19,B20} 
and $N$-fermion object like highly correlated electron-hole plasma. Such a plasma is spontaneously formed at low temperature through exciton condensation into electron-hole droplets as observed long ago in silicon or germanium \cite{droplet1, droplet2}. It can also be enforced through high pumping, the interacting exciton gas ultimately transforming into electron-hole plasma for density high enough to have exciton wave functions overlapping, in order to go beyond Mott transition.

Unabsorbed photons also produce fascinating effects in semiconductor materials which can appear as magic at first, due to the na\"{\i}ve idea that photons only act when they are absorbed. Indeed, while photons absorbed in semiconductor create real electron-hole pairs, photons which are not absorbed are yet coupled to semiconductor excitations which are then virtual. These virtual excitations interact in the same way as real ones, their interactions however stopping when the unabsorbed photons disappear, i.e., when the photon pulse is turned off. This is technologically quite nice because it allows to turn on and off interactions extremely fast. The effects induced by virtual excitations depend on photon detuning: as physically reasonable, the largest the effect, the closest the photons to resonance, i.e., to the possibility to create real excitations. Among effects induced by unabsorbed photons, we wish to cite a few that we find particularly impressive, namely, the exciton optical Stark effect \cite{Stark,Stark.bis} and the precession \cite{prec} or teleportation \cite{telep} of electron spin under irradiation with unabsorbed photon pulses.

(i) The problem of just \emph{one} exciton, made of one conduction electron and one valence hole, is quite simple and under complete control. Exciton being very similar to hydrogen atom, its energy spectrum made of bound and extended states is analytically known \cite{H1,H2} for 3D and exact 2D systems (i.e., zero-width quantum wells) if we forget interband Coulomb interaction \cite{Leuen}, inappropriately called ``electron-hole exchange'': while the exciton energy does not depend on carrier spins if we only consider intraband Coulomb processes, this \emph{interband} interaction pushes ``bright'' excitons with total spin $S_z=(\pm 1,0)$, i.e., excitons coupled to $(\sigma_{\pm},\pi)$ photons, above ``dark'' excitons having total spin $S_z=(\pm 2)$. As a direct quite important consequence, Bose-Einstein condensation of excitons has to occur in these dark states \cite{Leuen,BEC}, even if bright excitons are the ones which are created through photon absorption: indeed, dark excitons are formed in a natural way by carrier exchanges between two opposite-spin bright excitons, as nicely revealed by the Shiva diagrams which visualize carrier exchanges in the many-body theory for composite excitons we have recently constructed \cite{Comb3,Comb4}. 

Turning to photon absorption, we know that ground state excitons appear as a bright narrow peak in semiconductor spectrum. The coupling betwen one photon and one bound exciton, characterized by the so-called ``exciton oscillator strength'' $f_\mathrm{X}$, is indeed quite good because one plane wave photon with momentum $\v Q_p$ transforms into one plane wave bound exciton with same center-of-mass momentum. Coupling to ground state S exciton turns the largest because this state has the largest wave function for ``electron on top of hole'', carriers being possibly visualized as created this way by photon absorption.

(ii) The trion problem is far more complicated: trions are similar \cite{H2} to hydrogen ions $H^-$ or $H_2^+$ depending if these are made of two electrons plus one hole or two holes plus one electron. The existence of trion in semiconductors has ben predicted long ago by Lampert  \cite{T0,T1,T2,T3}, but evidenced recently only \cite{T4,T5}. The first difficulty to evidence trions comes from its extremely weak binding energy in bulk samples: the ``second'' electron is attracted not by a hole as for exciton, but by an electron-hole pair. Attraction by a dipole being much weaker than by an elementary charge, this makes the trion binding much weaker than exciton binding. Fortunately, reduction of space dimensionality strongly increases all bindings, as already seen from the fact that the exciton binding goes from $R_\mathrm{X}$ for 3D to $(4R_\mathrm{X})$ for 2D and infinity for 1D, this divergence being of course cut by the finite width of the quantum wire \cite{B7-6,H0,Xfil}. The possibility to now produce clean quantum wells allows to observe a line linked to trion well separated from the exciton line in quasi 2D geometry \cite{T5,T6,T7,T8,T9,T10,T11,T15,T17bis,T18,T19}.

From a mathematical point of view, the three-body problem does not have analytical solution, so that, as for hydrogen ion \cite{calculH+}, all precise calculations rely on heavy numerics, even for the ground state \cite{T2,T3}. Trion eigenstates split into fully bound composite particle (e-e-h), partially bound (e-h)+e, or fully dissociated but highly correlated carriers e+e+h. However, when speaking of trion, most people usually have in mind the fully bound (e-e-h) state.

We showed \cite{T12,T13,T16} that the physically relevant way to approach trion is to see it as a composite exciton interacting with one electron, the exciton being possibly dissociated into electron and hole, to also cover fully dissociated trion. This representation led us to introduce the Fourier transform ``in the exciton sense'' of the trion wave function which turns out to be the relevant quantity for trion oscillator strength $f_\mathrm{T}$. Using it, we showed \cite{T14} that $f_\mathrm{T}$ is one trion volume divided by one sample volume smaller than the exciton value $f_\mathrm{X}$. This makes trion impossible to evidence by photon absorption when formed out of a \emph{single} electron --- except in very poor samples, the sample volume being physically replaced by the coherence volume. 

Experimentally, lines associated to trion are seen in doped materials having a rather dense electron gas. It is of importance to understand that, in the absence of many-body effects, an increase of the electron density does not really help to increase the trion peak intensity at a given wavelength, because electrons being fermions, they all have different energies, so that photons producing a ground state trion out of these electrons must also have different energies. An increase of electron density thus tends to simply spread the trion line. The trion peak which is experimentally seen actually results from singular many-body effects.

Due to electron indistinguishability, the proper image of a trion in an electron gas is not the one of a 3-fermion object, but more the one of a complex $N$-fermion object, namely, an exciton made of a $(\pm 1/2)$ electron dressed by a cloud of $(-\pm 1/2)$ electrons, this exciton moreover exchanging its electron with  the $(\pm 1/2)$ electron cloud --- a quite complex many-body object definitely! To support this idea, we experimentally showed \cite{T17} that, indeed, the so-called trion line observed in doped quantum wells does  not come from ``a''  trion as commonly said, but results from many-body effects having similarities with Fermi-edge singularities \cite{Fesing}. Our old work on this problem however is not relevant because we dropped the electron-electron repulsion from the very first line of the calculations, while this repulsion partly cancels the electron-hole attraction, the balance between the two ruling the way the Fermi sea reacts to the \emph{sudden appearance of a photocreated exciton} --- not a photocreated hole as for standard Fermi edge singularities in metals. The many-body theory for composite excitons \cite{Comb4} we have recently constructed, in which the Pauli exclusion principle producing tricky carrier exchanges, is treated exactly, should help to tackle this quite complicated many-body problem. 

(iii) Biexciton has similarity with hydrogen molecule \cite{H2}, with fully bound (e-e-h-h) states and partially or fully dissociated highly correlated carriers like (e-h)+(e-h), (e-h)+e+h, or even e+e+h+h. The possibility that two excitons bind into a molecule was initially proposed by Lampert \cite{T1,B1}. The attraction of two dipoles is however very small. Moreover, in indirect gap semiconductors like Si and Ge, which were the materials mostly studied in the 60's, the biexciton line  falls under the very broad line associated with exciton condensation into electron-hole droplets. This is why biexcitonic molecules \cite{B2,B3} were first identified not in Ge or Si but in materials like CuCl \cite{B4}, Cu$_2$O \cite{B5}, AgBr \cite{B7bis,B7ter}, GaAs \cite{B10,B12}. However, by applying uniaxial stress on Si \cite{B7-4} or Ge \cite{B6}, it is possible to reduce the stability of the electron-hole plasma since this stability mostly comes from the multivalley degeneracy of the conduction band: with one valence band only, the exciton energy is lower than the plasma energy, so that excitonic molecules can show up as a peak below the exciton line.

From a mathematical point of view, the biexciton problem is a four-fermion problem, even more complicated than the trion three-fermion problem \cite{B7,B8,B9,B11,B14,B15,B16,B19}. By seeing biexciton as two interacting excitons, we could at first think that, since attraction between two dipoles is weaker than attraction between one dipole and one elementary charge, the biexciton problem should be simpler than the trion problem. However, in both cases, we face bound states; consequently, interactions, even far weaker than the bare Coulomb attraction between opposite single charges, must be treated \emph{exactly} in order to possibly generate the associated poles.

The unique but real conceptual advantage of biexciton compared to trion is the fact that, being made of an even number of fermions, excitons and biexcitons both have a bosonic nature, while electrons and trions are both fermion-like. Consequently, at the lowest order in the interactions, biexcitons can be formed out of excitons all having essentially the same energy. Biexcitons can also be piled up all at the same energy. In photoexcited semiconductors, this leads to a linear increase of the biexciton line intensity resulting from the absorption of one photon in the presence of a large number of excitons, in contrast with the trion peak in doped materials which would spread in the presence of a large number of electrons and thus needs singular many-body effects to be explained.

We will show that, as for trion, the ``bare'' oscillator strength for a fully bound biexciton made out of one exciton plus one photon is extremely small, being one biexciton volume divided by one sample volume smaller than the exciton oscillator strength. As a direct consequence,  the observed biexciton absorption line which is seen in photoexcited semiconductors \cite{Marcia1,Marcia2} results from a kind of bosonic enhancement --- an eleborate way to say that biexciton is formed out of one among $N_\mathrm{X}$ identical excitons. Luminescence experiments \cite{B7-5,B11bis,B16bis,B20} are somewhat more complex because when transforming one biexciton into one exciton plus one photon, bosonic enhancement can exist for both, the biexciton gas from which the biexciton which emits the photon originates, and the exciton gas, usually present in these luminescence experiments, which then increases its number by one unity. 

In the present work, we concentrate on one photon plus one exciton transforming into \emph{one} biexciton, and derive its ``bare'' oscillator strength. In our sense, this bare quantity should be the one called ``oscillator strength''. It of course enters the absorption or emission line amplitudes observed experimentally, but should be distinguished from them. These amplitudes usually contain ``external'' enhancements which can hide the intrinsic difficulty to form a complex quantum object through photon absorption. This is why, to call these amplitudes ``oscillator strengths'' as often done, does not seem to us physically appropriate.

Since the extremely small value of the biexciton oscillator strength could na\"{\i}vely lead to the conclusion that biexciton should not be seen, we end this work by a brief discussion of photon \emph{absorption} in the presence of an exciton gas, at the lowest order in many-body effects.  We also briefly comment on biexciton formation through two-photon absorption \cite{ B5bis,B7-4bis,B11,B11ter} and the misleadingly called ``giant'' biexciton oscillator strength \cite{B5bis} --- since, of course, this giant size is not intrinsic, but plainly comes from the increase of photon number available to form biexciton when increasing the laser intensity.

In order to properly describe one biexciton as two interacting composite excitons, we first have to determine the physically relevant set of spatial coordinates for two electrons located at $(\v r_e,\v r_{e'})$ and two holes located at $(\v r_h,\v r_{h'})$. Since \emph{we here concentrate on ground state molecular, i.e., fully bound biexciton}, we restrict to two electrons with opposite spins; and similarly for the two holes. We moreover consider that these excitons are located in quasi-2D type I quantum well \cite{B15bis}, in order to possibly consider hole spins $(\pm 3/2)$ only, for simplicity. All momenta are then in-plane momenta.

By turning to second quantization, as convenient for further study of many-body effects with biexcitons, we discuss in details the link between the symmetry properties of the biexciton wave function with respect to $(\v r_e,\v r_{e'})$ and $(\v r_h,\v r_{h'})$ and the decomposition of one biexciton into two composite excitons. This leads us to introduce the Fourier transform ``in the exciton sense'' of the biexciton wave function, similar to the one which appeared as highly relevant for trion \cite{T13,T14,T16}.
By using this Fourier transform, it becomes easy to derive the biexciton oscillator strength for the formation of one biexciton out of one photon plus a single exciton, and mostly to understand the reason for the drastic reduction of this bare oscillator strength compared to the exciton one. 

We wish to make  a last important comment. Most people improperly use the words ``exciton'', ``trion'' and ``biexciton'' for carriers trapped in quantum dots. This tends to shade the conceptual difference between carriers in dots and carriers in bulk, well or wire samples, i.e., samples having one free direction, at least. In the latter case, the distance between carriers in bound states results from a competition between kinetic energy which tends to spread out the carriers over the whole sample, and Coulomb attraction, this interaction having to be treated exactly in order to get the discrete bound states. By contrast, in dots, the distance between carriers is imposed by confinement. Coulomb energy then is a ``small'' perturbation with respect to confinement energy --- even if the absolute value of the Coulomb term is large, carriers in quantum dots being, by definition of dots, closer than the exciton Bohr radius. It then becomes easy to understand why photon absorption in neutral or charged quantum dot leads to very similar oscillator strengths: the additional carriers necessary to form ``dot trion'' or ``dot biexciton'', are already trapped in the dot, so that they are close to the photoexcited electron-hole pair. On the contrary, in bulk, well or wire, these additional carriers are ``free'', i.e., delocalized over the sample, so that in the process of forming fully bound trion or biexciton, we have to localize the additional carriers close to the photoexcited exciton. This brings a factor of the order of the bound object volume divided by sample volume, in the oscillator strength. Let us however note that, in poor samples having small coherence length, the additional carriers are already rather localized, making the coupling to photons easier, so that we can end with experimentally measured trion or biexciton oscillator strength very similar to the exciton one. In this work, we consider photon absorption in the presence of \emph{free} exciton, not electron-hole pair trapped in dot.

\section{Biexciton molecular ground state}

We consider two sets of electrons with mass $m_e$ and spin $\pm1/2$ and two sets of holes with mass $m_h$ and angular momentum $\pm 3/2$. Out of them, we can form two bright excitons $\pm1$ and two dark excitons $\pm 2$. For simplicity, we shall forget interband Coulomb processes, so that bright and dark excitons will be considered as degenerate.

The trion molecular ground state is known to have its two electrons $(+1/2,-1/2)$ in a singlet state --- as necessary for the ground state orbital part to be fully symmetrical. In the same way, the biexciton molecular ground state is made of a singlet state for electrons and a singlet-like state for holes. This leads us to expand the creation operator for molecular biexciton with center-of-mass momentum $\v K$ as
\begin{equation}
\maj B_{\v K}^\dag=\sum_{\v k_e,\v k_{e'},\v k_h,\v k_{h'}}\Psi^{(\v K)}_{\v k_e\v k_{e'}\v k_h\v k_{h'}}
\left(\frac{a_{\v k_e+}^\dag a_{\v k_{e'}-}^\dag-a_{\v k_e-}^\dag a_{\v k_{e'}+}^\dag}{\sqrt{2}}\right)
\left(\frac{b_{\v k_h+}^\dag b_{\v k_{h'}-}^\dag-b_{\v k_h-}^\dag b_{\v k_{h'}+}^\dag}{\sqrt{2}}\right)\ ,
\end{equation}
with $\Psi^{(\v K)}_{\v k_e\v k_{e'}\v k_h\v k_{h'}}\neq 0$ for $\v k_e+\v k_{e'}+\v k_h+\v k_{h'}=\v K$ only. Operator $a_{\v k\pm}^\dag$ creates an electron with orbital momentum $\v k$ and spin $\pm 1/2$, while, for quantum wells, $b_{\v k\pm}^\dag$ creates a hole with orbital momentum $\v k$ and angular momentum $\pm 3/2$.

By noting that $a_{\v k_e-}^\dag a_{\v k_{e'}+}^\dag=-a_{\v k_{e'}+}^\dag a_{\v k_e-}^\dag$, it is possible to rewrite Eq.(1) in a more compact form as
\begin{equation}
\maj B_{\v K}^\dag=\sum_{\v k_e,\v k_{e'},\v k_h,\v k_{h'}}\psi^{(\v K)}_{\v k_e\v k_{e'}\v k_h\v k_{h'}}
a_{\v k_e+}^\dag a_{\v k_{e'}-}^\dag b_{\v k_h+}^\dag b_{\v k_{h'}-}^\dag\ ,
\end{equation}
where $\psi$ is linked to $\Psi$ through
\begin{equation}
\psi^{(\v K)}_{\v k_e\v k_{e'}\v k_h\v k_{h'}}=\frac{1}{2}\left(\Psi^{(\v K)}_{\v k_e\v k_{e'}\v k_h\v k_{h'}}+\Psi^{(\v K)}_{\v k_{e'}\v k_{e}\v k_h\v k_{h'}}+\Psi^{(\v K)}_{\v k_e\v k_{e'}\v k_{h'}\v k_{h}}+\Psi^{(\v K)}_{\v k_{e'}\v k_{e}\v k_{h'}\v k_{h}}\right)\ ,
\end{equation}
so that the function $\psi^{(\v K)}_{\v k_e\v k_{e'}\v k_h\v k_{h'}}$ is a fully symmetrical function with respect to permutations $(\v k_e\leftrightarrow \v k_{e'})$ and $(\v k_h\leftrightarrow \v k_{h'})$,
\begin{equation}
\psi^{(\v K)}_{\v k_e\v k_{e'}\v k_h\v k_{h'}}=\psi^{(\v K)}_{\v k_{e'}\v k_{e}\v k_h\v k_{h'}}=\psi^{(\v K)}_{\v k_e\v k_{e'}\v k_{h'}\v k_{h}}\ .
\end{equation}

Note that, starting from Eq.(2), it is possible to write $\maj B_{\v K}^\dag$ as in Eq.(1) but with a symmetrical prefactor, namely, $(1/2)\psi^{(\v K)}_{\v k_e\v k_{e'}\v k_h\v k_{h'}}$ instead of
$\Psi^{(\v K)}_{\v k_e\v k_{e'}\v k_h\v k_{h'}} $. This expression of $\maj B_{\v K}^\dag$ would then also contain four terms which, due to Eq.(4), are in fact equal, so that this expression is not any better than the compact form of $\maj B_{\v K}^\dag$ given in Eq.(2), even if the singlet symmetry for electrons and for holes appears as more transparent in Eq.(1).

To get the wave function of the biexciton state $\maj B_{\v K}^\dag|v\rangle$ in terms of the prefactors $\psi^{(\v K)}$ appearing in the operator expansion, we use
\begin{equation}
\langle\v r_e,\v r_{e'}|a_{\v k_es}^\dag a_{\v k_{e'}s'}^\dag|v\rangle=\frac{1}{\sqrt{2}}\left(
\frac{e^{i\v k_e.\v r_e}}{\sqrt{L^D}}\frac{e^{i\v k_{e'}.\v r_{e'}}}{\sqrt{L^D}}|s,s'\rangle-
\frac{e^{i\v k_{e'}.\v r_e}}{\sqrt{L^D}}\frac{e^{i\v k_{e}.\v r_{e'}}}{\sqrt{L^D}}|s',s\rangle\right)\ ,
\end{equation}
where $|v\rangle$ is the electron-hole vacuum and $L^D$ the sample volume, with $D=2$ in 2D quantum well. From the symmetry property of $\psi^{(\v K)}_{\v k_e\v k_{e'}\v k_h\v k_{h'}}$ given in Eq.(4), we then find
\begin{equation}
\langle\v r_h,\v r_{h'},\v r_e,\v r_{e'}|\maj B_{\v K}^\dag|v\rangle=\psi^{(\v K)}(\v r_e,\v r_{e'},\v r_h,\v r_{h'})\left(\frac{|+-\rangle_e-|-+\rangle_e}{\sqrt{2}}\right)\left(\frac{|+-\rangle_h-|-+\rangle_h}{\sqrt{2}}\right)\ ,
\end{equation}
where $\psi^{(\v K)}(\v r_e,\v r_{e'},\v r_h,\v r_{h'})$ is just the Fourier transform of $\psi^{(\v K)}_{\v k_e\v k_{e'}\v k_h\v k_{h'}}$, namely,
\begin{equation}
\psi^{(\v K)}(\v r_e,\v r_{e'},\v r_h,\v r_{h'})=\sum_{\v k_e,\v k_{e'},\v k_h,\v k_{h'}}\psi^{(\v K)}_{\v k_e\v k_{e'}\v k_h\v k_{h'}}\langle\v r_e|\v k_e\rangle\langle\v r_{e'}|\v k_{e'}\rangle\langle\v r_h|\v k_h\rangle
\langle\v r_{h'}|\v k_{h'}\rangle\ ,
\end{equation}
with $\langle\v r|\v k\rangle=e^{i\v k.\v r}/L^{D/2}$.
Due to Eq.(4), this wave function is fully symmetrical with respect to permutations $(e\leftrightarrow e')$ and $(h\leftrightarrow h')$, as expected for ground state,
\begin{equation}
\psi^{(\v K)}(\v r_e,\v r_{e'},\v r_h,\v r_{h'})=\psi^{(\v K)}(\v r_{e'},\v r_{e},\v r_h,\v r_{h'})=\psi^{(\v K)}(\v r_e,\v r_{e'},\v r_{h'},\v r_{h})\ .
\end{equation}
It is however clear that these $(\v r_e,\v r_{e'},\v r_h,\v r_{h'})$ coordinates are not the physically relevant ones to describe a molecular state. Indeed, one of these coordinates must be the center-of-mass position. In the next section, we are going to discuss what could be the other three coordinates.

\section{Relevant coordinates for one biexciton}

\subsection{Spatial coordinates}

We look for $\v R_i$, with $i=(1,2,3,4)$, in terms of $(\v r_e,\v r_{e'},\v r_h,\v r_{h'})$ and we take as granted that the center-of-mass position of these four carriers,
\begin{equation}
\v R_\mathrm{BX}=\frac{m_e\v r_e+m_e\v r_{e'}+m_h\v r_h+m_h\v r_{h'}}{2m_e+2m_h}\ ,
\end{equation}
has to be one of the relevant biexciton spatial coordinates, namely, $\v R_4=\v R_{\mathrm{BX}}$. We are then left with determining the other three coordinates as $\v R_i=x_i\v r_e+x_i'\v r_{e'}+y_i\v r_h+y_i'\v r_{h'}$ with $i=(1,2,3)$. This corresponds to $3\times 3=9$ unknowns if we restrict to normalized $\v R_i$'s. For sure, the physically relevant set must be such that $[\v R_i,\v P_j]=i\hbar\delta_{ij}$, where $\v P_j$ is the momentum operator associated to coordinate $\v R_j$. These commutation relations are necessary to have decoupling between these new coordinates, i.e., absence of non-diagonal terms $\v P_i.\v P_j$ with $i\neq j$ in the resulting Hamiltonian. While this decoupling condition is enough to fully determine the relative motion coordinate $\v r=\v r_e-\v r_h$ in the case of one exciton, it is not enough for more complex objects such as trion or biexciton. Indeed, in the case of biexciton, conditions $[\v R_i,\v P_j]=0$ for $i\neq j$ lead to 6 conditions for 9 unknowns, while conditions $[\v R_i,\v P_i]=i\hbar$ just allow us to get the associated effective masses. To fully determine the other three spatial coordinates, we thus have to bring some additional ideas.

Having in mind to represent biexciton as two interacting excitons, we are led to choose for two of these three additional coordinates, the electron-hole distances in the two excitons, for instance, $\v r_e-\v r_h$ and $\v r_{e'}-\v r_{h'}$. The third coordinate, which now is fully determined by the decoupling condition, then appears to be the distance between the two centers of mass of the excitons made of $(e,h)$ and $(e',h')$. It is however clear that we could as well choose $\v r_{e'}-\v r_h$ and $\v r_e-\v r_{h'}$, the third coordinate then being the distance between the two centers of mass of the excitons made of $(e',h)$ and $(e,h')$. Since the two electrons have opposite spins as well as the two holes, the excitons resulting from this exchange have different spins. Let us call $(\v r_e,\v r_h)$ the coordinates of the 1/2 electron and 3/2 hole,and $(\v r_{e'},\v r_{h'})$ the coordinates of the -1/2 electron and -3/2 hole. The physically relevant sets of coordinates then read as
\begin{eqnarray}
\v r_1&=&\v r_{e'}-\v r_h\ ,\nonumber\\
\v r_{-1}&=&\v r_e-\v r_{h'}\ ,\nonumber\\
\v u_1&=&\frac{m_e\v r_{e'}+m_h\v r_h}{m_e+m_h}-\frac{m_e\v r_e+m_h\v r_{h'}}{m_e+m_h}\ ,
\end{eqnarray}
for bright excitons having total spin $S=\pm1$, i.e., excitons made of $(\mp1/2,\pm3/2)$ carriers (see Fig.1(a)), and
\begin{eqnarray}
\v r_2&=&\v r_{e}-\v r_h\ ,\nonumber\\
\v r_{-2}&=&\v r_{e'}-\v r_{h'}\ ,\nonumber\\
\v u_2&=&\frac{m_e\v r_{e}+m_h\v r_h}{m_e+m_h}-\frac{m_e\v r_{e'}+m_h\v r_{h'}}{m_e+m_h}\ ,
\end{eqnarray}
for dark excitons having total spin $S=\pm 2$, made of $(\pm1/2,\pm3/2)$ carriers, (see Fig.1(b)).

In terms of these new coordinates, the biexciton wave function $\psi^{(\v K)}(\v r_e,\v r_{e'},\v r_h,\v r_{h'})$ appears as
\begin{equation}
\psi^{(\v K)}(\v r_e,\v r_{e'},\v r_h,\v r_{h'})=\frac{e^{i\v K.\v R_\mathrm{BX}}}{\sqrt{L^D}}\,\phi(\v r_1,\v r_{-1},\v u_1)\ ,
\end{equation}
the parity condition, Eq.(8), now reading as
\begin{equation}
\phi(\v r_1,\v r_{-1},\v u_1)=\phi(\v r_2,\v r_{-2},\v u_2)=\phi(\v r_{-1},\v r_1,-\v u_1)\ .
\end{equation}

For completeness, let us note that the biexciton Hamiltonian which, written in terms of $(\v r_e,\v r_{e'},\v r_h,\v r_{h'})$ coordinates, reads as
\begin{eqnarray}
H_\mathrm{BX}=\frac{\v p_e^2+\v p_{e'}^2}{2m_e}+\frac{\v p_h^2+\v p_{h'}^2}{2m_h}+V(\v r_e-\v r_{e'})
+V(\v r_h-\v r_{h'})\hspace{2cm}\nonumber\\
-V(\v r_e-\v r_h)-V(\v r_e-\v r_{h'})-V(\v r_{e'}-\v r_h)-V(\v r_{e'}-\v r_{h'})\ ,
\end{eqnarray}
where $V(\v r)=e^2/r$, transforms, in terms of $(\v R_\mathrm{BX},\v r_1,\v r_{-1},\v u_1)$ coordinates, as
\begin{equation}
H_\mathrm{BX}=\frac{\v P_\mathrm{BX}^2}{4(m_e+m_h)}+h(\v r_1,\v r_{-1},\v u_1)\ .
\end{equation}
$h(\v r_1,\v r_{-1},\v u_1)$, which can be seen as the biexciton relative motion Hamiltonian, can be possibly replaced by
$h(\v r_2,\v r_{-2},\v u_2)$ due to the $(\v r_e\leftrightarrow \v r_{e'})$ invariance of $H_\mathrm{BX}$. Its precise value reads
\begin{equation}
h(\v r_1,\v r_{-1},\v u_1)=h_1+h_{-1}+\frac{\v p_{\v u_1}^2}{2\mu_\mathrm{BX}}+\mathcal{V}(\v r_1,\v r_{-1},\v u_1)\ .
\end{equation}
$h_1=\v p_1^2/2\mu_\mathrm{X}-V(\v r_1)$ is the relative motion Hamiltonian for exciton made of electron $(-1/2)$ located at $\v r_{e'}$ and hole $(+3/2)$ located at $\v r_h$, and similarly for $h_{-1}$, the exciton relative motion mass being $\mu_\mathrm{X}^{-1}=m_e^{-1}+m_h^{-1}$. The biexciton relative motion mass $\mu_\mathrm{BX}$ appearing in Eq.(16) is such that
\begin{equation}
\frac{1}{\mu_\mathrm{BX}}=\frac{1}{M_\mathrm{X}}+\frac{1}{M_\mathrm{X}}=\frac{2}{M_\mathrm{X}}\ ,
\end{equation}
where $M_\mathrm{X}=m_e+m_h$: it corresponds to the mass associated to the relative motion of the two exciton centers of mass. The coupling between these two excitons is insured by the Coulomb potential between their carriers. For bright excitons made from $(e,h')$ and $(e',h)$, it reads
\begin{eqnarray}
\mathcal{V}(\v r_1,\v r_{-1},\v u_1)&=&V(\v r_e-\v r_{e'})+V(\v r_h-\v r_{h'})-V(\v r_e-\v r_h)-V(\v r_{e'}-\v r_{h'})\nonumber\\
&=& \frac{e^2}{|\gamma_h(\v r_1-\v r_{-1})+\v u_1|}+ \frac{e^2}{|\gamma_e(\v r_1-\v r_{-1})-\v u_1|}\nonumber\\
& &\ \ \  - \frac{e^2}{|\gamma_e\v r_1+\gamma_h\v r_{-1}-\v u_1|}- \frac{e^2}{|\gamma_h\v r_1+\gamma_e\v r_{-1}+\v u_1|}\ ,
\end{eqnarray}
where $\gamma_e=1-\gamma_h=m_e/(m_e+m_h)$.

\subsection{Momentum coordinates}

(i) Let $(\v k_e,\v k_{e'},\v k_h,\v k_{h'})$ be the momentum coordinates associated to 
$(\v r_e,\v r_{e'},\v r_h,\v r_{h'})$ and $(\v K,\v k_1,\v k_{-1},\v Q_1)$ the momentum coordinates associated to $(\v R_\mathrm{BX},\v r_1,\v r_{-1},\v u_1)$, the biexciton center-of-mass momentum being such that $\v K=\v k_e+\v k_{e'}+\v k_h+\v k_{h'}$. Since the two excitons have equal masses, we are led to split the center-of-mass momentum $\v K$ equally between the two bright exciton center-of-mass momenta which then read as $(\v Q_1+\v K/2)$ and $(-\v Q_1+\v K/2)$. These momenta are ultimately split between the corresponding carriers, according also to their masses, so that we end with
\begin{eqnarray}
\v k_{e'}&=&\v k_1+\gamma_e\left(\v Q_1+\frac{\v K}{2}\right)\ ,\nonumber\\
\v k_{h}&=&-\v k_1+\gamma_h\left(\v Q_1+\frac{\v K}{2}\right)\ ,\nonumber\\
\v k_{e}&=&\v k_{-1}+\gamma_e\left(-\v Q_1+\frac{\v K}{2}\right)\ ,\nonumber\\
\v k_{h'}&=&-\v k_{-1}+\gamma_h\left(-\v Q_1+\frac{\v K}{2}\right)\ .
\end{eqnarray}
$\v k_1$ and $\v k_{-1}$ are the relative motion momenta of the bright excitons having spins ($+1$) and ($-1$), while $\v Q_1$ is the momentum for the relative motion of these two $(+1,-1)$ excitons. From Eqs.(10,19), it is easy to show the following relation
\begin{equation}
\v k_e.\v r_e+\v k_{e'}.\v r_{e'}+\v k_h.\v r_h+\v k_{h'}.\v r_{h'}=\v K.\v R_\mathrm{BX}+\v k_1.\v r_1+
\v k_{-1}.\v r_{-1}+\v Q_1.\v u_1\ .
\end{equation}

Within these relevant spatial and momentum coordinates, the wave function in momentum space, defined as the Fourier transform of the wave function in $\v r$ space, is found to read as
\begin{equation}
\phi_{\v k_1,\v k_{-1},\v Q_1}=\int d\v r_1\,d\v r_{-1}\,d\v u_1\,\langle\v k_1|\v r_1\rangle\langle\v k_{-1}|\v r_{-1}\rangle\langle\v Q_1|\v u_1\rangle \phi(\v r_1,\v r_{-1},\v u_1)\ ,
\end{equation}
with $\phi(\v r_1,\v r_{-1},\v u_1)$ defined by Eq.(12). By using Eqs.(7,12,19,20), we end with a Fourier transform $\phi_{\v k_1,\v k_{-1},\v Q_1}$ of $\phi(\v r_1,\v r_{-1},\v u_1)$ which is nothing but
\begin{equation}
\phi_{\v k_1,\v k_{-1},\v Q_1}\equiv\psi^{(\v K)}_{\v k_e,\v k_{e'},\v k_h,\v k_{h'}}\ .
\end{equation}
$(\v k_e,\v k_{e'},\v k_h,\v k_{h'})$,  in the RHS of the above equation, read in terms of $(\v k_1,\v k_{-1},\v Q_1)$ according to Eq.(19): it is indeed possible to check, using Eq.(21), that the RHS of the above equation is $\v K$-independent.

(ii) In the same way, the spatial coordinates $(\v R_\mathrm{BX},\v r_2,\v r_{-2},\v u_2)$ for biexcitons seen as two interacting dark excitons, are associated to momenta $(\v K,\v k_2,\v k_{-2},\v Q_2)$ defined by the following relations
\begin{eqnarray}
\v k_{e}&=&\v k_2+\gamma_e\left(\v Q_2+\frac{\v K}{2}\right)\ ,\nonumber\\
\v k_{h}&=&-\v k_2+\gamma_h\left(\v Q_2+\frac{\v K}{2}\right)\ ,\nonumber\\
\v k_{e'}&=&\v k_{-2}+\gamma_e\left(-\v Q_2+\frac{\v K}{2}\right)\ ,\nonumber\\
\v k_{h'}&=&-\v k_{-2}+\gamma_h\left(-\v Q_2+\frac{\v K}{2}\right)\ ,
\end{eqnarray}
for which we also have
\begin{equation}
\v k_e.\v r_e+\v k_{e'}.\v r_{e'}+\v k_h.\v r_h+\v k_{h'}.\v r_{h'}=\v K.\v R_\mathrm{BX}+\v k_2.\v r_2+
\v k_{-2}.\v r_{-2}+\v Q_2.\v u_2\ .
\end{equation}

\section{Biexciton creation operator}

We now have all the tools to rewrite the biexciton creation operator given in Eq.(2) in terms of exciton operators, as convenient to derive physical quantities dealing with biexciton such as the oscillator strength. For that, we use the link between excitons and free carriers, namely,
\begin{equation}
B_{i,S}^\dag=\sum_{\v k_e,\v k_h}a_{\v k_e,s}^\dag b_{\v k_h,m}^\dag\,\langle\v k_h,\v k_e|i\rangle\ ,
\end{equation}
\begin{equation}
a_{\v k_e,s}^\dag b_{\v k_h,m}^\dag=\sum_iB_{i,S}^\dag\,\langle i|\v k_e,\v k_h\rangle\ ,
\end{equation}
where $i=(\nu_i,\v Q_i)$, with $\nu_i$ being the $i$ exciton relative motion index for bound or extended state and $\v Q_i$ the $i$ exciton center-of-mass momentum; $\langle\v k_h,\v k_e|i\rangle$ is the $i$ exciton wave function in momentum space. The spin variable $S$ is such that $S=s+m$, the link between exciton spin and carrier spins being one to one in the case of quantum well. These equations can also be written in terms of the exciton relative motion wave function in momentum space $\langle\v k|\nu_i\rangle$ as
\begin{equation}
B_{\nu_i,\v Q_i,S}^\dag=\sum_{\v k}a_{\v k+\gamma_e\v Q_i,s}^\dag b_{-\v k+\gamma_h\v Q_i,m}^\dag
\langle \v k|\nu_i\rangle\ ,
\end{equation}
\begin{equation}
a_{\v k_e,s}^\dag b_{\v k_h,m}^\dag=\sum_{\nu}B_{\nu,\v k_e+\v k_h,S}^\dag\,\langle\nu|\gamma_h\v k_e-\gamma_e\v k_h\rangle\ .
\end{equation}

If, in Eq.(2), we associate the free carrier creation operators as $(a_{\v k_{e'}-}^\dag b_{\v k_h+}^\dag)
(a_{\v k_e+}^\dag b_{\v k_{h'}-}^\dag)$ and then use Eq.(28), we get the biexciton creation operator $\maj B_{\v K}^\dag$ as a sum of products of two bright-exciton operators. Using Eq.(19), we are led to write $\maj B_{\v K}^\dag$ as
\begin{equation}
\maj B_{\v K}^\dag=\sum_{\nu_1,\nu_{-1},\v Q_1}\phi_{\nu_1,\nu_{-1},\v Q_1}\,B_{\nu_1,\v Q_1+\v K/2,+1}^\dag
B_{\nu_{-1},-\v Q_1+\v K/2,-1}^\dag\ ,
\end{equation}
where the prefactor of this decomposition of one biexciton into two bright excitons is given, due to Eq.(22), by
\begin{equation}
\phi_{\nu_1,\nu_{-1},\v Q_1}=\sum_{\v k_1,\v k_{-1}}\langle\nu_1|\v k_1\rangle\langle\nu_{-1}|\v k_{-1}\rangle\,\phi_{\v k_1,\v k_{-1},\v Q_1}\ .
\end{equation}
According to Eq.(21), the above equation also reads
\begin{equation}
\phi_{\nu_1,\nu_{-1},\v Q_1}=\int d\v r_1\,d\v r_{-1}\,d\v u_1\,\langle\nu_1|\v r_1\rangle\langle\nu_{-1}|
\v r_{-1}\rangle\langle\v Q_1|\v u_1\rangle\phi(\v r_1,\v r_{-1},\v u_1)\ ,
\end{equation}
where $\langle\v r|\nu_i\rangle$ is the relative motion wave function of exciton $i$ in $\v r$ space.
This $\phi_{\nu_1,\nu_{-1},\v Q_1}$ prefactor appears as a ``mixed Fourier transform'' of the biexciton wave function written within the physically relevant spatial coordinates, namely, the distances between electron and hole in the two bright excitons $(\v r_1,\v r_{-1})$ and the distance $\v u_1$ between the centers of mass of these two bright excitons. Indeed, Eq.(31) is a bare Fourier transform with respect to $\v u_1$, but a Fourier transform ``in the exciton sense'' with respect to the other two coordinates $\v r_1$ and $\v r_{-1}$, since these coordinates are transformed into exciton indices $\nu_1$ and $\nu_{-1}$. A similar mixed Fourier transform already appeared when we described a trion as one exciton interacting with one electron \cite{T13,T14,T16}. 

Equations (29,31), along with the expression of the biexciton oscillator strength given below in Eq.(39), are the main results of the present paper.

Note that, due to possible carrier exchanges, we could as well write $\maj B_{\v K}^\dag$ in terms of two dark excitons, these excitons being made of electron-hole pairs $(\v k_e,+1/2)$, $(\v k_h,+3/2)$ and $(\v k_{e'},-1/2)$, $(\v k_{h'},-3/2)$. However, as seen below, the decomposition of biexciton in terms of bright excitons is the appropriate one for physical effects involving photons.

\section{Biexciton oscillator strength}

\subsection{Formal expression of the biexciton oscillator strength}

A very direct way to reach the biexciton oscillator strength is to consider a sample already having one circularly polarized exciton, for instance $(\nu_0,\v Q_\mathrm{X},-1)$ and to calculate the absorption of one photon with momentum $\v Q_p$, frequency $\omega_p$ and opposite circular polarization. By using $(x+i\epsilon)^{-1}=\mathcal{P}(1/x)-i\pi\delta(x)$, the absorption for initial state $|I\rangle$, given by the Fermi golden rule,
\begin{equation}
A(\omega_p)=2\pi\,\sum_F\left|\langle F|W^\dag|I\rangle\right|^2\delta(\mathcal{E}_F-\mathcal{E}_I)\ ,
\end{equation}
reads in terms of the imaginary part of the response function
\begin{equation}
S(\omega_p)=\langle I|W\frac{1}{\mathcal{E}_I-H+i\epsilon}W^\dag|I\rangle\ ;
\end{equation}
If we only keep resonant terms, the semiconductor coupling $W^\dag$, associated to the absorption of a  $\v Q_p$ photon with $\sigma_+$ polarization and $\alpha_{\v Q_p,+}^\dag$ creation operator, can be reduced to
\begin{eqnarray}
W^\dag&=&\Omega^\ast\,\alpha_{\v Q_p,+}\,\sum_{\v k}a_{\v k+\gamma_e\v Q_p,-}^\dag b_{-\v k+\gamma_h\v Q_p,+}^\dag\nonumber\\
&=&\alpha_{\v Q_p,+}\,\sum_\nu\Omega_\nu^\ast B_{\nu,\v Q_p,+1}^\dag\ ,
\end{eqnarray}
where $\Omega^\ast$ is the free-pair Rabi coupling while $\Omega_\nu^\ast=\Omega^\ast\sqrt{L^D}\langle\nu|\v r=\v 0\rangle$ is the Rabi coupling to exciton $\nu$, as easy to recover from Eq.(28). Let us stress that, by choosing to also quantize the photon field, the semiconductor-photon interaction becomes time-independent, so that we do not have to use the rotating frame to eliminate time.

The initial state $|I\rangle$ relevant for the formation of one biexciton out of one exciton with spin $(-1)$ and one photon with $\sigma_+$ polarization, is
\begin{equation}
|I\rangle=B_{\nu_0,\v Q_\mathrm{X},-1}^\dag|v\rangle\otimes \alpha_{\v Q_p,+}^\dag
|0\rangle\ ,
\end{equation}
where $|0\rangle$ is the photon vacuum.The energy of this initial state is $\mathcal{E}_I=E_{\nu_0,\v Q_\mathrm{X}}+\omega_p$. The state $W^\dag|I\rangle$ appearing in the response function (33) has no more photon, but two electron-hole pairs. The simplest way to derive the absorption associated to the formation of a molecular biexciton with center-of-mass momentum $\v K$, is to inject in front of $(\mathcal{E}_I-H+i\epsilon)^{-1}$ in Eq.(33), the closure relation for the Hamiltonian $H$ two-pair eigenstates, and, in this closure relation, to only keep the molecular biexciton state, namely, $\maj B_{\v K}^\dag|v\rangle\otimes |0\rangle$. This gives the part of the response function associated to the formation of a molecular biexciton as
\begin{eqnarray}
S_\mathrm{BX}^{(\v K)}&=&\langle I|W\frac{1}{\mathcal{E}_I-H+i\epsilon}\maj B_{\v K}^\dag|v\rangle\otimes|0\rangle\langle 0|\otimes\langle v|\maj B_{\v K}W^\dag|I\rangle\nonumber \\
&=&\frac{f_\mathrm{BX}^{(\v K)}}{\omega_p+E_{\nu_0,\v Q_\mathrm{X}}-\maj E_{\v K}+i\epsilon}\ .
\end{eqnarray}
Equation (34) readily gives $f_\mathrm{BX}^{(\v K)}$ as
\begin{equation}
f_\mathrm{BX}^{(\v K)}=\left|\langle 0|\otimes\langle v|\maj B_{\v K}W^\dag|I\rangle\right|^2=\left|\langle v|\maj B_{\v K}
\sum_\nu\Omega_\nu B_{\nu,\v Q_p,+1}^\dag B_{\nu_0,\v Q_\mathrm{X},-1}^\dag|v\rangle\right|^2\ .
\end{equation}
If we now write the biexciton creation operator according to Eq.(29) and note that $\langle v|B_{i',S'}
B_{i,S}^\dag|v\rangle=\delta_{i',i}\delta_{S',S}$, we see that the scalar product appearing in Eq.(37) imposes $\nu_1=\nu$ and $\nu_{-1}=\nu_0$ as well as $\v Q_1+\v K/2=\v Q_p$ and $-\v Q_1+\v K/2=\v Q_\mathrm{X}$, so that $f_\mathrm{BX}^{(\v K)}$ splits as $f_\mathrm{BX}^{(\v K)}=\delta_{\v K,\v Q_p+\v Q_\mathrm{X}}\,f_\mathrm{BX}$, as physically expected since the photocreated biexciton center of mass must have the momentum of the photon-exciton pair from which it is constructed. The amplitude of the biexciton oscillator strength then appears as
\begin{equation}
f_\mathrm{BX}=\left|\sum_\nu L^{D/2}\Omega\langle\nu|\v r=\v 0\rangle\phi_{\nu_1=\nu,\nu_{-1}=\nu_0,\v Q_1=(\v Q_p-\v Q_\mathrm{X})/2}^\ast\right|^2\ .
\end{equation}
By using Eq.(31) for $\phi_{\nu_1,\nu_{-1},\v Q_1}$, it is possible to perform the summation over $\nu$ through closure relation. The biexciton oscillator strength $f_\mathrm{BX}$, for a biexciton made out of one exciton $\nu_0$ with momentum $\v Q_\mathrm{X}$ and spin $(-1)$ and one $\sigma_+$ photon with momentum $\v Q_p$, ultimately reads in terms of the biexciton relative motion wave function in real space $\phi(\v r_1,\v r_{-1},\v u_1)$ defined in Eq.(12) as
\begin{eqnarray}
f_\mathrm{BX}&=&|\Omega|^2L^D\left|\int d\v r_{-1}\,d\v u_1\,\langle\frac{\v Q_p-\v Q_\mathrm{X}}{2}|\v u_1\rangle\langle\nu_0|\v r_{-1}\rangle\phi(\v r_1=\v 0,\v r_{-1},\v u_1)\right|^2\nonumber\\
&=&|\Omega|^2L^D\left|\phi(\v r_1=\v 0,\nu_0,(\v Q_p-\v Q_\mathrm{X})/2)\right|^2\ .
\end{eqnarray}

The above result can be physically understood by saying that, as in the case of exciton for which the oscillator strength,
\begin{equation}
f_\mathrm{X}=|\Omega|^2L^D|\varphi_\mathrm{X}(\v r=\v 0)|^2\ ,
\end{equation}
makes appear the relative motion wave function $\varphi_\mathrm{X}(\v r)$ of the photocreated exciton with ``electron on top of hole'', through $\v r=\v 0$, the part of the biexciton wave function which corresponds to the exciton created by the $\sigma_+$ photon absorption, has also to appear with ``electron on top of hole'', through $\v r_1=\v 0$. The second exciton $\nu_0$, already present in the sample, enters the oscillator strength through its Fourier transform ``in the exciton sense'', via $\langle\nu_0|\v r_{-1}\rangle$. This amounts to replace the spatial coordinate $\v r_{-1}$ of the (-1) exciton, in the biexciton wave function $\phi(\v r_1,\v r_{-1},u_1)$, by the relative motion index $\nu_0$ of this exciton. Finally, the localization of the center of mass of the second exciton, initially delocalized through a $\v Q_\mathrm{X}$ center-of-mass plane wave, in the vicinity of the photocreated exciton, as necessary to form a molecular fully bound biexcitonic state, is enforced through a standard Fourier transform via $\langle\frac{\v Q_p-\v Q_\mathrm{X}}{2}|\v u_1\rangle$. This amounts to replace the biexciton relative motion coordinate $\v u_1$, in the biexciton wave function $\phi(\v r_1,\v r_{-1},u_1)$, by the momentum $(\v Q_p-\v Q_\mathrm{X})/2$ associated to this coordinate.

This oscillator strength has similarity with the one of a trion made of one $\v Q_p$ exciton and one $\v k_e$ free electron. Indeed, the trion oscillator strength reads as \cite{T14}
\begin{eqnarray}
f_\mathrm{T}&=&|\Omega|^2L^D\left|\int d\v u\,\langle\beta_e\v Q_p-\beta_\mathrm{X}\v k_e|\v u\rangle\,\phi_\mathrm{T}(\v r=\v 0,\v u)\right|^2\nonumber\\
&=&|\Omega|^2L^D\left|\phi_\mathrm{T}(\v r=\v 0,\beta_e\v Q_p-\beta_\mathrm{X}\v k_e)\right|^2\ ,
\end{eqnarray}
where $\beta_e=1-\beta_\mathrm{X}=m_e/(2m_e+m_h)$, while $\v u=\v R_\mathrm{X}-\v r_e$ is the distance between the photocreated exciton center of mass and the electron already present. Note that, in the case of biexciton, the photocreated exciton has the same mass as the exciton present in the sample, so that $\beta_e$ in Eq.(41) has to be replaced by $(m_e+m_h)/(2m_e+2m_h)=1/2$, so that $\beta_\mathrm{X}$ is also equal to 1/2, in agreement with the $\v Q$ prefactors in Eq.(39).

\subsection{Comparison between exciton, trion and biexciton oscillator strengths}

In this last section, we use the above results to qualitatively estimate and compare the oscillator strengths of exciton, trion and biexciton. The two latter ones are going to be found far smaller than the exciton oscillator strength for a quite fundamental reason: the quantum particle already present in the sample, which is completely delocalized over the whole sample (except for quantum dot), ends by being localized close to the photocreated exciton in order to form the bound state of interest. This localization, for sure costly, leads to a strong reduction of the coupling to photon. 

Let us now recover this quite simple understanding from the expressions of oscillator strengths given above in Eqs.(39,40,41).

(i) Due to dimensional arguments, the \emph{normalized} wave function for the relative motion of a bound state exciton extending over a Bohr Radius $a_\mathrm{X}$ is such that $|\varphi(\v r=\v 0)|^2\propto a_\mathrm{X}^{-D}$ within an irrelevant numerical factor. Consequently, in 2D, Eq.(40) gives
\begin{equation}
f_\mathrm{X}\propto|\Omega|^2\left(\frac{L}{a_\mathrm{X}}\right)^2\ ,
\end{equation}
the sample size $L$ having to be replaced by the coherence length $L^\ast$ in real experimental situations: indeed, this $L$ factor can be traced back to the center-of-mass part $\langle\v R|\v Q\rangle$ of the exciton wave function, which has been taken as $e^{i\v Q.\v R}/L^{D/2}$. In practice, this plane wave extends over the coherence length $L^\ast$ only.

(ii) In the same way, the normalized wave function for the relative motion of a trion made of one exciton extending over $r\simeq a_\mathrm{X}$ and one electron localized at a ``trion Bohr radius'' $a_\mathrm{T}$ from this exciton, $u\simeq a_\mathrm{T}$, must be such that $|\phi_\mathrm{T}(\v r=\v 0,\v u=\v 0))|^2\propto (a_\mathrm{X}a_\mathrm{T})^{-D}$. We then note that $\langle\v u|\v k\rangle=e^{i\v k.
\v u}/L^{D/2}$ while the $\phi_\mathrm{T}(\v r,\v u)$ trion relative motion wave function forces $u$ to stay of the order of $a_\mathrm{T}$. For $(Q_p,k_e)$ small compared to $a_\mathrm{T}^{-1}$, we then find that $\langle\beta_e\v Q_p-
\beta_\mathrm{X}\v k_e|\v u\rangle$ in Eq.(41) stays essentially equal to $1/L$ for $D=2$, so that integration over $\v u$ in this Eq.(41) brings a $a_\mathrm{T}^2$ factor. All this leads to
\begin{eqnarray}
f_\mathrm{T}&\propto& |\Omega|^2L^2\left(a_\mathrm{T}^2\frac{1}{L}\frac{1}{a_\mathrm{T}a_\mathrm{X}}\right)^2=|\Omega|^2
\left(\frac{a_\mathrm{T}}{a_\mathrm{X}}\right)^2\nonumber\\
&\propto&\left(\frac{a_\mathrm{T}}{L}\right)^2f_\mathrm{X}\ ,
\end{eqnarray}
in agreement with the result we previously obtained in ref.\cite{T14}.

(iii) We now use the same procedure for biexciton. The extension of the relative motion of a fully bound molecular biexciton is of the order of $a_\mathrm{X}$ with respect to the two exciton coordinates $(r_1,r_{-1})$ but of the order of $a_\mathrm{BX}$ for the $u_1$ coordinate, where $a_\mathrm{BX}$ is the biexciton spatial extension. This gives $|\phi(\v r_1=\v 0,\v r_{-1}=\v 0,\v u_1=\v 0)|^2\propto (a_\mathrm{X}^2a_\mathrm{BX})^{-D}$. For $D=2$ and $(Q_p,Q_\mathrm{X})$ small compared to $a_\mathrm{BX}^{-1}$, the factor $\langle\v Q|\v u_1\rangle$ in Eq.(39) provides a factor $1/L$, while integration over $\v u_1$ in this equation brings a factor $a_\mathrm{BX}^2$. Since $\langle\nu_0|\v r\rangle\propto
a_\mathrm{X}^{-D/2}$, while integration over $\v r_{-1}$ brings a factor $a_\mathrm{X}^2$, we end with
\begin{eqnarray}
f_\mathrm{BX}&\propto&|\Omega|^2L^2\left(a_\mathrm{X}^2a_\mathrm{BX}^2\frac{1}{L}\frac{1}{a_\mathrm{X}}\frac{1}{a_\mathrm{X}^2a_\mathrm{BX}}\right)^2=|\Omega|^2\left(\frac{a_\mathrm{BX}}{a_\mathrm{X}}\right)^2\nonumber\\
&\propto&\left(\frac{a_\mathrm{BX}}{L}\right)^2\,f_\mathrm{X}\ .
\end{eqnarray}

The prefactors $(a_\mathrm{T}/L)^2$ or $(a_\mathrm{BX}/L)^2$ appearing in $f_\mathrm{T}$ or $f_\mathrm{BX}$ physically correspond to the localization of the additional quantum particle, either the electron or the exciton, originally delocalized over the whole volume $L^D$, within a trion or biexciton volume $a_\mathrm{T}^D$ or $a_\mathrm{BX}^D$ from the photocreated exciton (see Fig.2). The complete similarity between trion and biexciton is, on that respect, quite enlightening.

Note that, if instead of making a molecular biexciton, i.e., a quantum particle with its two excitons at roughly $a_\mathrm{BX}$ from each other, we form a biexciton dissociated into two excitons, the $u_1$ coordinate in $\phi(\v r_1,\v r_{-1},\v u_1)$ would extend over $L$ instead of staying at $a_\mathrm{BX}$; consequently, we would then have $|\phi(\v r_1=\v 0,\v r_{-1}=\v 0,\v u_1=\v 0)|^2
\propto (a_\mathrm{X}^2L)^{-D}$. The same argument would lead to an oscillator strength for this partly dissociated biexciton of the order of
\begin{equation}
f_\mathrm{XX}\propto|\Omega|^2L^2\left(a_\mathrm{X}^2L^2\frac{1}{L}\frac{1}{a_\mathrm{X}}\frac{1}{a_\mathrm{X}^2L}\right)^2=|\Omega|^2\left(\frac{L}{a_\mathrm{X}}\right)^2\ .
\end{equation}
This is nothing but the free-exciton oscillator strength $f_\mathrm{X}$, as expected since, in this case, the absorbed photon essentially adds one free exciton which weakly interacts with the exciton present in the sample.

The above discussion, essentially based on physical understanding, makes appear the ``biexciton extension'' $a_\mathrm{BX}$ as a key parameter. This has to be contrasted with previous works dedicated to quantitative understanding \cite{quant}. These works use variational procedure through trial wave function for biexciton with more standard spatial coordinates than the ones we use here. In particular, these coordinates do not fulfill $[\v R_i,\v P_j]=i\hbar\delta_{ij}$. In these approaches the biexciton extension $a_{BX}$, which characterizes the distance between the two exciton centers of mass, is not a natural parameter, so that direct comparison with the present work is not really possible.

\subsection{Biexciton in the presence of an exciton gas}

We now qualitatively explain why the creation of a biexciton in the presence of an exciton gas is quite different from the creation of a trion in the presence of an electron gas.

(i) Trions, made of an odd number of carriers, are composite fermions, so that they cannot be piled up all at the same energy. A simple way to see it, is to consider an initial state made of $N_e$ electrons at zero temperature, these electrons having momentum extending from $|\v k|=0$ to $|\v k|=k_F$. Trions which can be formed by absorption of one $(\omega_p,\v Q_p)$ photon in this Fermi sea (see Fig.3(a)), have  momenta $\v Q_\mathrm{T}=\v Q_p+\v k$ extending between $|\v Q_p|\simeq 0$ and $|\v Q_p|+k_F\simeq k_F$. Energy conservation in photon absorption imposes
\begin{equation}
\omega_p+\frac{\v k^2}{2m_e}=\mathcal{E}_\mathrm{T}+\frac{(\v Q_p+\v k)^2}{2(2m_e+m_h)}\ ,
\end{equation}
where $\mathcal{E}_\mathrm{T}$ is the ground state trion binding energy. Consequently, for photon momentum $|\v Q_p|\simeq 0$, absorption in the absence of many-body effects would extend between $\omega_p=\mathcal{E}_\mathrm{T}$ and 
and $\omega_p=\mathcal{E}_\mathrm{T}-\epsilon_F[1-m_e/(2m_e+m_h)]$, where $\epsilon_F=\v k_F^2/2m_e$ is the Fermi sea energy level (see Fig.3(b)). Since the weight of each individual line, i.e., the oscillator strength, is vanishingly small, such a broadening of the absorption spectrum does not help to see the bound state trion contribution. As previously explained \cite{T17}, the observed line actually results from singular many-body effects induced by the sudden creation of one virtual exciton in an electron sea (see Fig.3(c)). In other words, the other electrons do not stay spectators during photon absorption, as in Fig.3(a), but play a crucial singular role in the observed broad peak.

(ii) We now turn to the absorption of one photon in a gas of $N_\mathrm{X}$ excitons. Excitons being boson-like particles, these $N_\mathrm{X}$ excitons can be considered as all being in the same ground state $(\nu_0,\v Q_0\simeq \v 0)$ if, as a first approximation, we forget their interactions. In the linear response to a photon field, the absorbed photon can then form a molecular biexciton with any of these $N_\mathrm{X}$ ground state excitons (see Fig.4(a)). As a result, the absorption probability to form a biexciton is going to increase linearly with $N_\mathrm{X}$. This linear increase with exciton density (see Fig.4(b)), instead of spreading out as for the formation of trions, ends by producing a noticeable absorption line amplitude at the biexciton energy minus the exciton energy, even if the oscillator strength for biexciton made out of a single exciton is extremely small.

Another, more elaborate, way to present this understanding is to say that, before photon absorption, the initial state of the system can be taken as $B_0^{\dag N}|v\rangle$, where $B_0^\dag$ is the ground state exciton creation operator. Biexciton creation requires the destruction of one of these $N_\mathrm{X}$ excitons. This would simply bring a $\sqrt{N_\mathrm{X}}$ factor if excitons were taken as elementary bosons, i.e., if we forgot exciton interaction through the Pauli blocking between their fermionic components. Since the coupling to the initial state appears squared in the absorption, we recover a $N_\mathrm{X}$ increase.

Of course, here also, the other excitons do not stay spectators in the absorption process, so that this simple picture is going to be modified by many-body effects (see Fig.4(c)). The most important ones come, as usual, from fermion exchanges resulting from the exciton composite nature induced by the Pauli exclusion principle. One example of these fermion exchanges is shown in Fig.4(d). We must however stress that, in this diagram, two among the $N_\mathrm{X}$ excitons present in the sample, enter the exchange process, so that the effect this exchange induces is going to vary as $N_\mathrm{X}^2$, i.e., quadratically with exciton density. Interactions between excitons are also going to shift the biexciton line due to the fact that the average energy of the exciton making the biexciton changes with density. In addition, the excitons are going to interact with the biexciton to change its binding energy.
The precise study of all these coupled many-body effects between the photocreated exciton and a sea of excitons, as well as their consequence on the biexciton absorption line, is definitely far beyond the scope of the present work. 

We however wish to stress that, while many-body effects are definitely crucial to explain why a trion-related line is experimentally seen, these are somewhat marginal in the case of biexciton, a linear increase of the absorption line resulting from the bosonic nature of both, exciton and biexciton, already explaining why biexciton can be rather easily evidenced in the absorption spectrum of photoexcited semiconductors.

\subsection{Biexciton through two-photon absorption}

Although not considered here, let us end this work by some comments on a somewhat more standard procedure \cite{B5bis,B7-4bis,B11,B11ter} to form biexciton, namely, through two-photon absorption. Since two photons are needed, the absorption line increases quadratically with photon number --- which actually is the signature for two-photon process. By contrast, since exciton needs one photon only to be formed, the exciton line increases linearly with the number of available photons. This led Hanamura \cite{B5bis} to bring the idea of a ``giant biexciton oscillator strength''. This idea completely shades the physics of the problem. Indeed, when the laser intensity is large enough, the biexciton line amplitude which increases quadratically with laser intensity, can overcome the exciton line amplitude which only increases linearly. This is straightforward.

Nevertheless, the intrinsic coupling between two photons and one bound exciton still is very poor compared to the one between one photon and one bound exciton. Indeed, two photons which correspond to two plane waves, have to be transformed into one plane wave only, the one of the biexciton center of mass. By contrast, one plane wave photon transforms into one plane wave exciton in the case of bound exciton formation, which explains the good coupling between photon and bound exciton. It is just because a lot of photons are available in the case of a powerful laser pulse, that the biexciton line is indeed seen in a two-photon absorption. This clearly shows that associating absorption line amplitude to oscillator strength can lead to a profound misunderstanding of the microscopic physics involved.

It can be also of interest to note that this very simple approach of counting plane wave numbers before and after absorption allows us to understand why the coupling between two photons and bound biexciton is far larger than the coupling to the band. Indeed, in the latter case, we transform two plane waves into four plane waves (two for the two free electrons and two for the two free holes), while in the case of bound biexciton, we end with one plane wave only.

\section{Conclusion}

In this paper, we concentrate on the physical understanding of one biexciton, to control its coupling to photon. With that in mind, we first determine the physical set of spatial coordinates which, along with the center-of-mass position, allows the description of one biexciton as two interacting excitons. The prefactor of the biexciton creation operator, written as an expansion in products of exciton creation operators, is found to be the Fourier transform ``in the exciton sense'' of the biexciton wave function written within this physical set of spatial coordinates. This Fourier transform appears as the relevant quantity for the oscillator strength associated to the transformation of one photon plus one exciton into one biexciton. This oscillator strength is found to be one biexciton volume divided by one sample --- or coherence --- volume smaller than the exciton oscillator strength. Comparison between exciton, trion and biexciton oscillator strengths is also given to enlighten the physics behind the photon coupling to these bound states. Comments on biexciton formation through two-photon absorption are also given for completeness.

\vspace{1cm}

We wish to thank Marcia Portella-Oberli for inducing this work and for enlightening discusions. We also acknowledge for very many valuable comments from Jim Wolfe on biexciton absorption and luminescence.

\newpage

\begin{figure}[t]
\vspace{-3cm}
\centerline{\scalebox{0.7}{\includegraphics{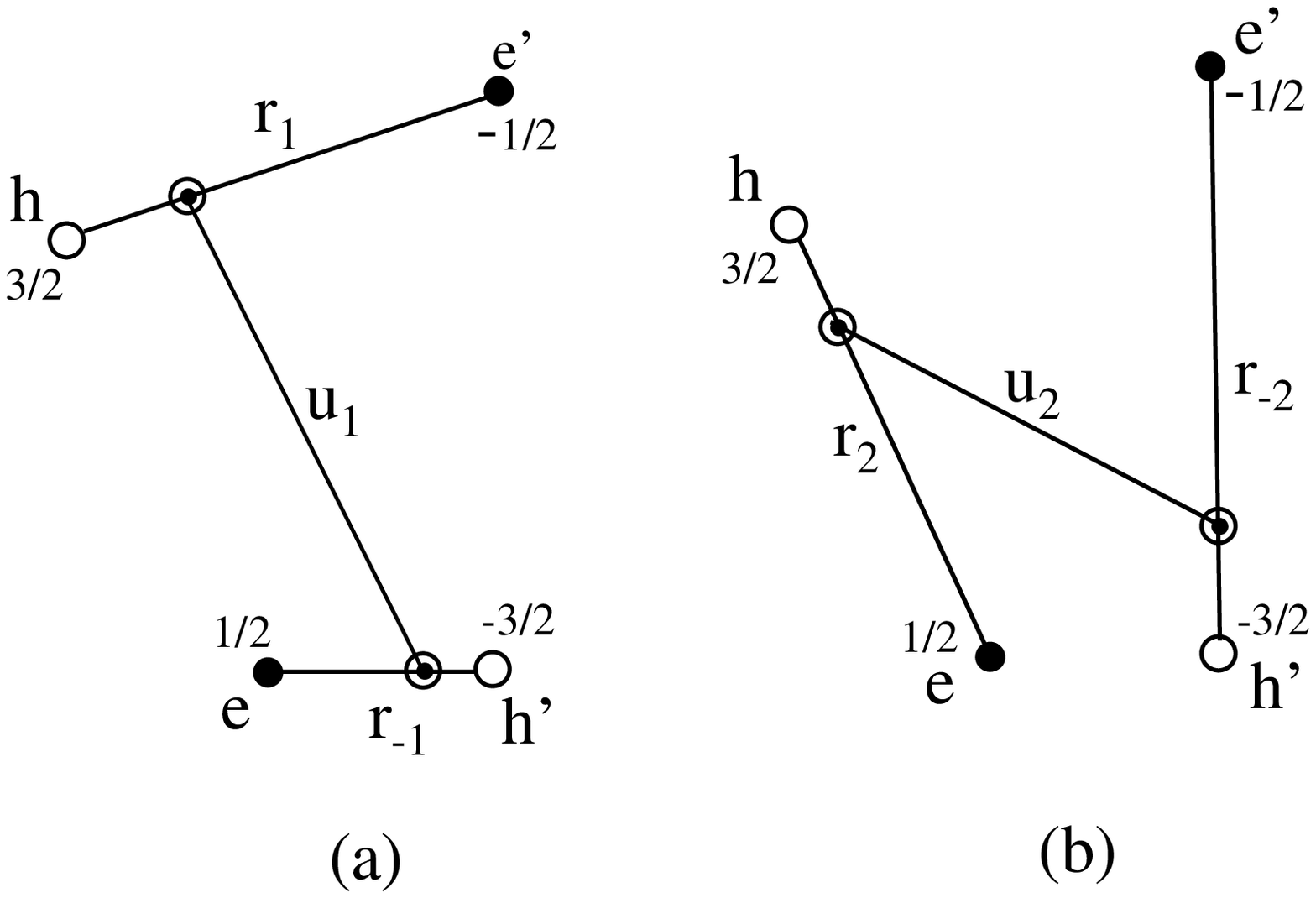}}}
\vspace{-5cm}
\caption{The physically relevant spatial coordinates for a ground state molecular biexciton made of $(\pm 1/2)$ electrons and $(\pm 3/2)$ holes, seen either as (a) two interacting bright excitons with total spins $(+1,-1)$ or (b) two interacting dark excitons with total spins $(+2,-2)$.}
\end{figure}

\clearpage

\begin{figure}[t]
\vspace{-2cm}
\centerline{\scalebox{0.7}{\includegraphics{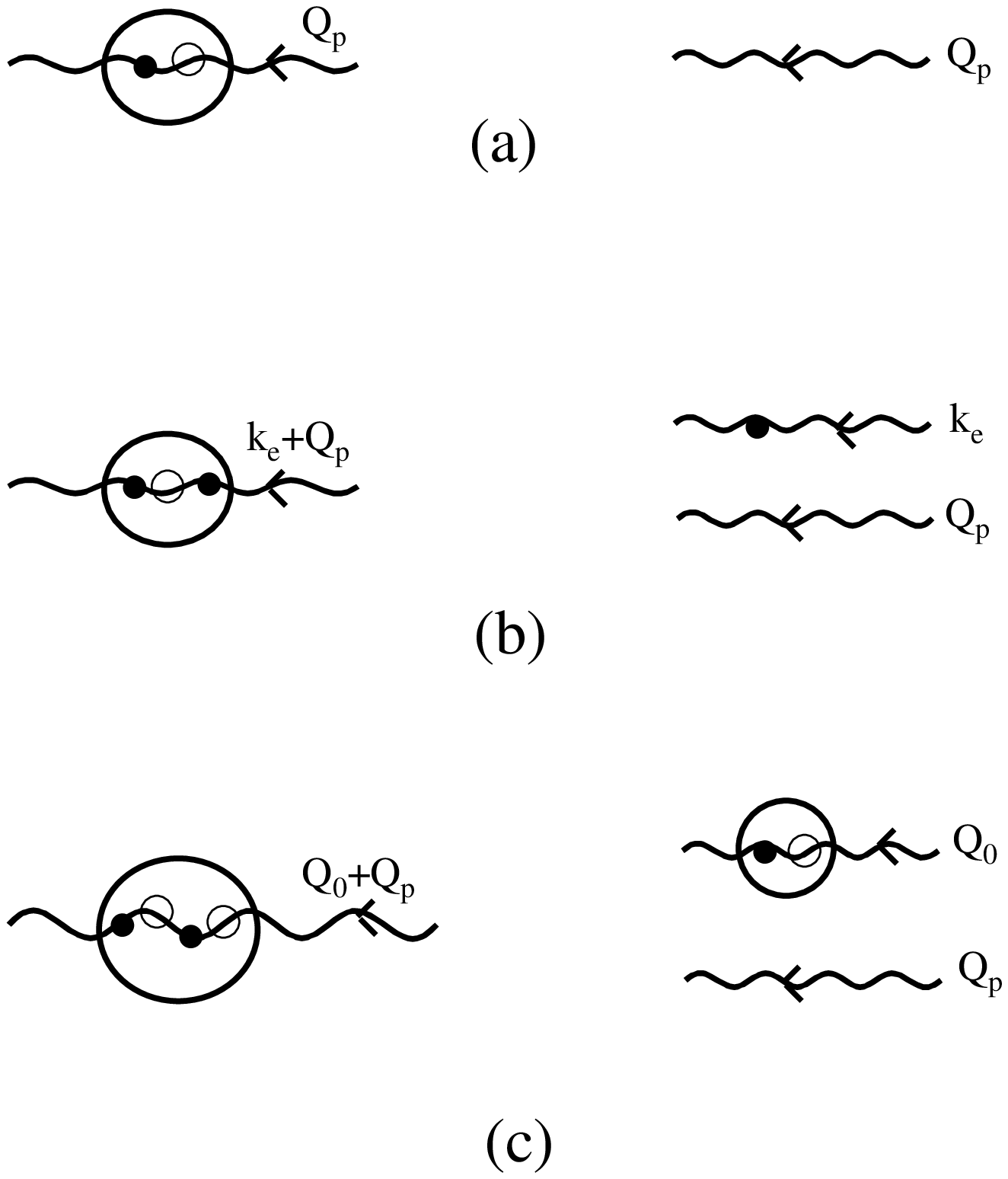}}}
\vspace{-5cm}
\caption{(a): The coupling between one photon $\v Q_p$ and one exciton with same center-of-mass momentum is quite good because one plane wave transforms into one plane wave. (b), (c): In contrast, two plane waves transform into one plane wave only when a molecular trion is formed by photon absorption out of a single electron $\v k_e$, as in (b), or a molecular biexciton is formed by photon absorption out of a single ground state exciton $\v Q_0$, as in (c). This makes the trion and biexciton oscillator strengths a bound particle volume divided by a sample (or coherence) volume smaller than the exciton oscillator strength.}
\end{figure}
 
\clearpage

\begin{figure}[t]
\centerline{\scalebox{0.7}{\includegraphics{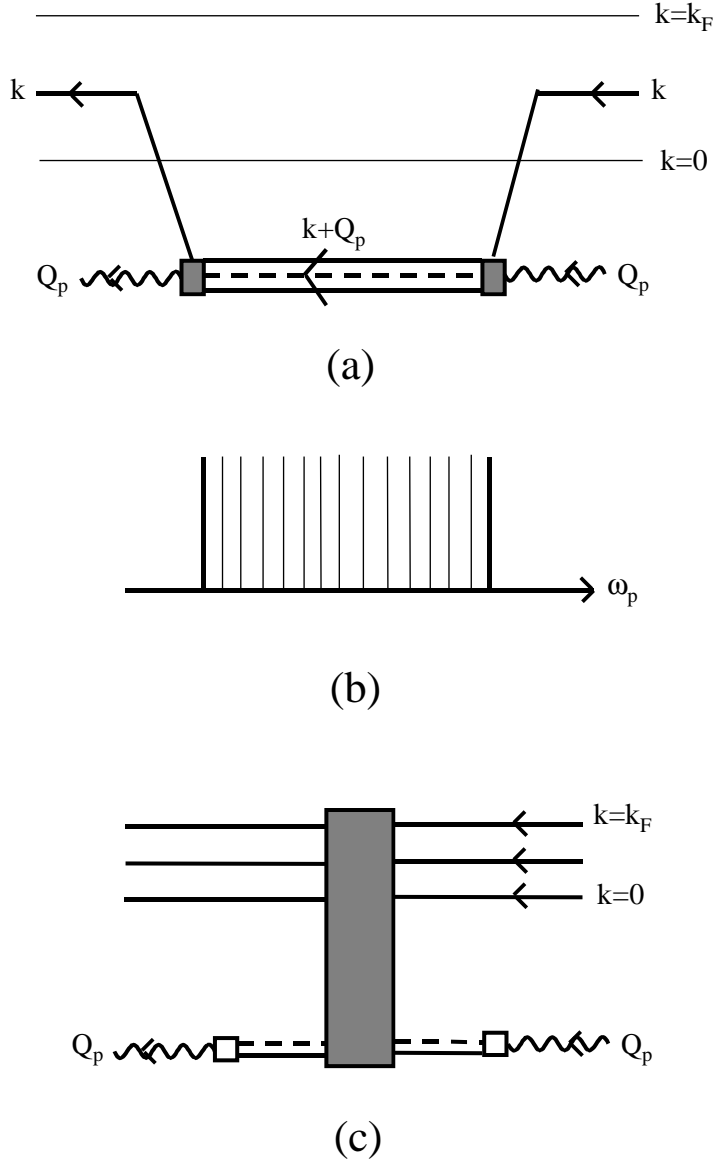}}}
\caption{(a): In the absence of interactions, a trion with momentum $\v k+\v Q_p$ is made out of one exciton coupled to a $\v Q_p$ photon and one $\v k$ free electron. (b): For electrons with momenta extending between 0 and $k_F$, the trions which can be formed lead to photon absorption between $\mathcal{E}_\mathrm{T}-\epsilon_F[1-m_e/(2m_e+m_h)]$ and $\mathcal{E}_\mathrm{T}$: the trion line is intrinsically broad but still vanishingly small in amplitude, in the absence of many-body effects. (c): The observed ``trion line'' actually results from singular many-body effects taking place ``inside the grey box'', between the photocreated exciton and the electron gas.}
\end{figure}

\clearpage

\begin{figure}[t]
\centerline{\scalebox{0.7}{\includegraphics{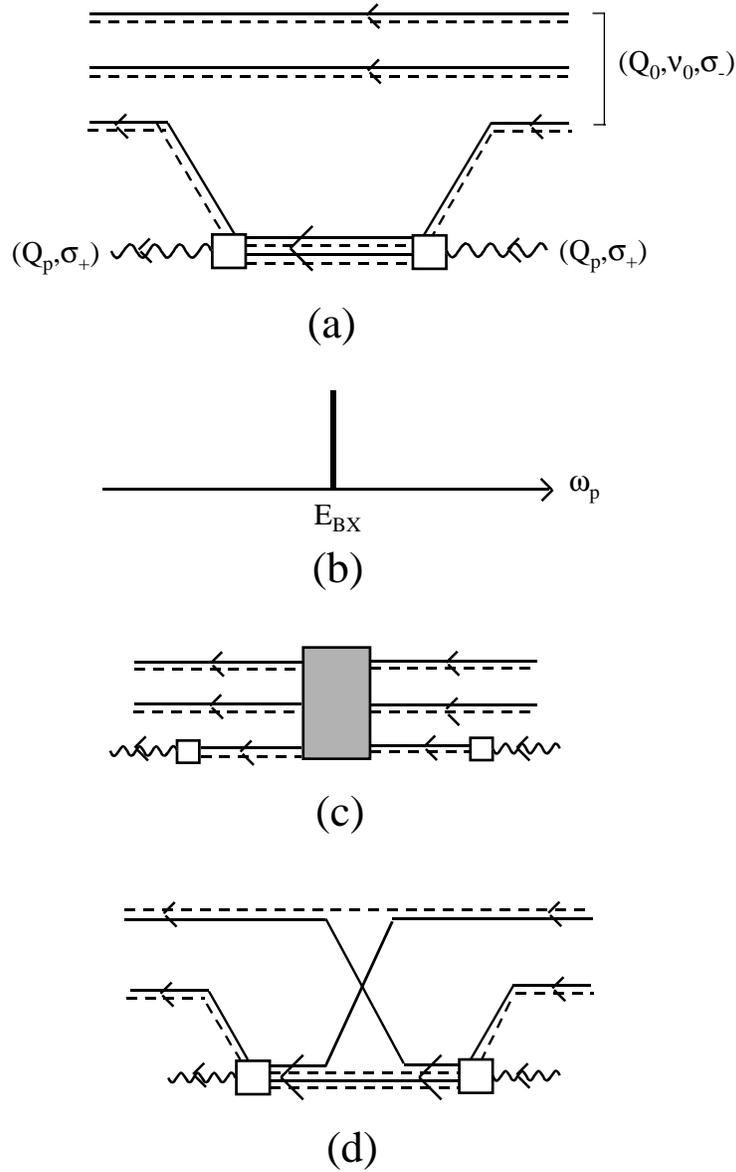}}}
\caption{(a): In the absence of interactions, one biexciton is formed out of an exciton coupled to a $(\v Q_p,\sigma_+)$ photon and one among $N_\mathrm{X}$ identical excitons $(\v Q_0,\nu_0,\sigma_-)$ already present in the sample. (b): The presence of very many equivalent excitons tends to compensate the poor coupling between a single exciton and one photon, through a linear increase of the biexciton line amplitude at the \emph{same} energy $E_\mathrm{BX}$. (c): Many-body effects can also take place ``in the grey box'' between the photocreated exciton and the exciton gas. (d): The dominant ones come from carrier exchanges, the biexcitons coupled to the ``in'' and ``out'' photons having a different electron (or a different hole). The effect shown in this figure, which needs two excitons among $N_\mathrm{X}$ to take place, has a quadratic dependence in exciton density.}
\end{figure}

\end{document}